\documentclass[11pt]{article}
\usepackage{amsmath}
\usepackage{amsfonts}
\usepackage[english]{babel}
\usepackage{amssymb}
\usepackage{amsthm}
\usepackage{graphicx}
\usepackage[a4paper]{geometry}
\usepackage{hyperref}
\usepackage{amstext}
\usepackage{caption}
\usepackage{etoolbox}
\usepackage{makeidx}
\usepackage{amsfonts}
\usepackage{authblk} 
\usepackage{sectsty}
\usepackage{amscd}
\usepackage{mathrsfs}
\usepackage{dsfont}
\usepackage{enumitem} 
\usepackage[]{latexsym}
\usepackage{braket}
\usepackage{caption}

\usepackage{xcolor}

\newcommand{\be}{\begin{equation}}
\newcommand{\ee}{\end{equation}}
\newcommand{\bea}{\begin{eqnarray}}
\newcommand{\eea}{\end{eqnarray}}

\newcommand{\stsp}{\mathcal{S}}

\newcommand{\lag}{\mathfrak{L}}

\newcommand{\gr}{\mathrm{g}}
\newcommand{\gri}{\mathrm{g}^{-1}}
\newcommand{\grd}{\dot{\mathrm{g}}}

\newtheorem*{proof*}{Proof}

\title{Hamilton-Jacobi theory and Information Geometry}
\author{Florio M. Ciaglia$^{1,2}$ and Fabio Di Cosmo$^{1,2}$  and Giuseppe Marmo$^{1,2}$ \\
$^{1}$  Dipartimento di Fisica, Universit\`a di Napoli ``Federico II",\\ Via Cinthia Edificio 6, I-80126 Napoli, Italy  \\ $^{2}$ INFN-Sezione di Napoli, Via Cinthia Edificio 6, I-80126 Napoli, Italy} 

\date{}

\begin{document}

\maketitle

\begin{abstract}
Recently, a method to dynamically define a divergence function $D$ for a given statistical manifold $(\mathcal{M}\,,g\,,T)$ by means of the Hamilton-Jacobi theory associated with a suitable Lagrangian function $\lag$ on $T\mathcal{M}$ has been proposed.
Here we will review this construction and lay the basis for an inverse problem where we assume the divergence function $D$ to be known and we look for a Lagrangian function $\lag$ for which $D$ is a complete solution of the associated Hamilton-Jacobi theory.
To apply these ideas to quantum systems, we have to replace probability distributions with probability amplitudes.
\end{abstract}

\section{Introduction}

In the field of information geometry, divergence functions are ubiquitous objects.
A divergence function $D$ is a positive semi-definite two-point function defined on $\mathcal{M}\times \mathcal{M}$, where $\mathcal{M}$ is the manifold underlying the statistical model $(\mathcal{M}\,,g\,,T)$ under study (see \cite{amari-information_geometry_and_its_application,a.a.v.v.-differential_geometry_in_statistical_inference,amari_nagaoka-methods_of_information_geometry}), such that $D(m_{1}\,,m_{2})=0$ if and only if $m_{1}=m_{2}$.
Roughly speaking, the value $D(m_{1}\,,m_{2})$ is interpreted as a ``measure of difference'' between the probability distributions parametrized by  $m_{1}$ and $m_{2}$.
The exact meaning of this difference depends on the explicit model considered.
If we imbed classical probabilities in the space of quantum systems, i.e., we replace probabilities with probability amplitudes, it is still possible to define divergence functions and derive metric tensors for quantum states.
For instance, when $\mathcal{M}=\mathcal{P}(\mathcal{H})$ is the space of pure states of a quantum system with Hilbert space $\mathcal{H}$, Wootter has shown (see \cite{wootters-statistical_distance_and_hilbert_space}) that a divergence function $D$ providing a meaningful notion of statistical distance between pure states may be introduced by means of the concepts of distinguishability  and statistical fluctuations in the outcomes of measurements.
It turns out that this statistical distance coincides with the Riemannian geodesic distance associated with the Fubini-Study metric on the complex projective space.
On the other hand, when $\mathcal{M}=\stsp_{n}$ is the manifold of positive probability measure on $\chi=\{1,...,n\}$, and $D$ is the Kullback-Leibler divergence function (see \cite{amari-information_geometry_and_its_application,a.a.v.v.-differential_geometry_in_statistical_inference,amari_nagaoka-methods_of_information_geometry}), then the meaning of the ``difference'' between $m_{1}$ and $m_{2}$ as measured by $D$ is related with the asymptotic estimation theory for an empirical probability distribution extracted from independent samples associated with a given probability distribution (see \cite{amari-information_geometry_and_its_application}).
One of the main features of a divergence function $D$ is the possibility to extract from it a metric tensor $g$, and a skewness tensor $T$ on $\mathcal{M}$ using an algorithm involving  iterated derivatives of $D$ and the restriction to the diagonal of $\mathcal{M}\times \mathcal{M}$ (see \cite{amari-information_geometry_and_its_application,a.a.v.v.-differential_geometry_in_statistical_inference,amari_nagaoka-methods_of_information_geometry}).
Given a statistical model $(\mathcal{M}\,,g\,,T)$ there is always a divergence function whose associated tensors are precisely $g$ and $T$ (see \cite{matumoto-any_statistical_manifold_has_a_contrast_function}), and, what is more, there is always an infinite number of such divergence functions.
In the context of classical information geometry, all statistical models share the ``same'' metric $g$, called the Fisher-Rao metric. 
This metric arise naturally when we consider $\mathcal{M}$ as immersed in the space $P(\chi)$ of probability distributions on the measure space $\chi$, and, provided some additional requirements on symmetries are satisfied, it is essentially unique (see \cite{amari_nagaoka-methods_of_information_geometry,cencov-statistical_decision_rules_and_optimal_inference}).
This means that, once the statistical manifold $\mathcal{M}\subset P(\chi)$ is chosen, all the admissible divergence functions must give back the Fisher-Rao metric $g$.
On the other hand, different admissible divergence functions lead to different third order symmetric tensors $T$.
Quite interestingly, the metric tensor $g$ is no longer unique in the quantum context (see \cite{petz-monotone_metrics_on_matrix_spaces}).

In a recent work (\cite{ciaglia_dicosmo_felice_mancini_marmo_perez-pardo-hamilton-jacobi_approach_to_potential_functions_in_information_geometry}), a dynamical approach to divergence functions has been proposed.
The main idea is to read a divergence function $D$, or more generally, a potential function for a given statistical model $(\mathcal{M}\,,g\,,T)$, as the Hamilton principal function  associated with a suitably defined Lagrangian function $\lag$ on $T\mathcal{M}$ by means of the Hamilton-Jacobi theory (see \cite{carinena_gracia_martinez_munoz-lecanda_roman-roy-geometric_hamilton-jacobi_theory,lanczos-the_variational_principles_of_mechanics}).
From this point of view, a divergence function $D$ becomes a dynamical object, that is, the function $D$ is no more thought of as some fixed kinematical function on the double of the manifold of the statistical model, but, rather, it becomes the Hamilton principal function associated with a Lagrangian dynamical system on the tangent bundle of the manifold of the statistical model.
In the variational formulation of dynamics \cite{lanczos-the_variational_principles_of_mechanics}, the solutions of the equations of motion are expressed as the critical points of the action functional:
\begin{equation}\label{eqn: action}
I\left(\gamma\right)=\int_{t_{\mathrm{in}}}^{t_{\mathrm{fin}}}\,\lag\left(\gamma\,,\dot{\gamma}\right)\,\mathrm{d}t\,,
\end{equation}
where $\gamma$ are curves on $\mathcal{M}$ with fixed extreme points $m(t_{\mathrm{in}})=m_{\mathrm{in}}$ and $m(t_{\mathrm{fin}})=m_{\mathrm{fin}}$, and $\lag$ is the Lagrangian function of the system.
In order to avoid technical details, we will always assume that $\lag$ is a regular Lagrangian (see \cite{marmo_ferrario_lovecchio_morandi_rubano-the_inverse_problem_in_the_calculus_of_variations_and_the_geometry_of_the_tangent_bundle}).
The evaluation of the action functional on a critical point $\gamma_{c}$ gives a two-point function\footnote{In general, this function depends on the additional parameters $t_{\mathrm{in}}$ and $t_{\mathrm{fin}}$, however we will always take $t_{\mathrm{in}}=0$ and $t_{\mathrm{fin}}=1$.}:
\begin{equation}\label{eqn: H-J solution}
S\left(m_{\mathrm{in}}\,,m_{\mathrm{fin}}\right)=I(\gamma_{c})\,,
\end{equation}
which is known in the literature as the Hamilton principal function.
When a given dynamics admits of alternative Lagrangian description, it is possible to integrate alternative Lagrangians along the same integral curves and get different  potential  functions.
If the determinant of the matrix of the mixed partial derivatives of $S$ is different from zero, then it is possible to prove (see \cite{lanczos-the_variational_principles_of_mechanics}) that $S$ is a complete solution of the Hamilton-Jacobi equation for the dynamics:
\begin{equation}
H\left(x\,,\frac{\partial S}{\partial x}\,,t\right) + \frac{\partial S}{\partial t}=0\,,
\end{equation}
where $H$ is the Hamiltonian function (\cite{carinena_ibort_marmo_morandi-geometry_from_dynamics_classical_and_quantum}) associated with the Lagrangian $\lag$.
In this case, $S(m_{\mathrm{in}}\, , m_{\mathrm{fin}})$ is called a complete solution for the Hamilton-Jacobi theory.
It turns out that the existence of a complete solution $S$ forces the dynamical system associated with the Lagrangian function $\lag$ to be completely integrable, that is, to adimit $n=dim(\mathcal{M})$ functionally independent constants of the motion which are transversal to the fibre of $T\mathcal{M}$ (see \cite{carinena_gracia_martinez_munoz-lecanda_roman-roy-geometric_hamilton-jacobi_theory}).
The main result of \cite{ciaglia_dicosmo_felice_mancini_marmo_perez-pardo-hamilton-jacobi_approach_to_potential_functions_in_information_geometry} is to prove that, given any  statistical model $(\mathcal{M}\,,g\,,T)$, the Lagrangian functions:
\be
\lag_{\alpha}=\dfrac{1}{2} g_{jk}(x)v^jv^k + \dfrac{\alpha}{6} T_{jkl}(x)v^jv^kv^l \,, 
\label{Lagrangian}
\ee  
labelled by the one-dimensional real parameter $\alpha$, are  such that their associated Hamilton principal functions are potential functions for $(\mathcal{M}\,,g\,,T)$ in the sense that they allow to recover $g$ and $T$ as follows:
\be
\left.\frac{\partial^{2}\,S_{\alpha}}{\partial x^{j}_{\mathrm{fin}}\partial x^{k}_{\mathrm{in}}}\right|_{x_{\mathrm{in}}=x_{\mathrm{fin}}}=-g_{jk}(x)\,,
\label{metric2}
\ee
\be
\left. \dfrac{\partial ^3 S_{\alpha}}{\partial x^{l}_{\mathrm{in}} \partial x^{k}_{\mathrm{in}} \partial x^{j}_{\mathrm{fin}} }\right|_{x_{\mathrm{in}}=x_{\mathrm{fin}}} - \left. \dfrac{\partial ^3 S_{\alpha}}{\partial x^{l}_{\mathrm{fin}} \partial x^{k}_{\mathrm{fin}} \partial x^{j}_{\mathrm{in}} }\right|_{x_{\mathrm{in}}=x_{\mathrm{fin}}} = 2\alpha T_{jkl}(x)\,.
\label{tensor2}
\ee
The functions $S_{\alpha}$ are not in general fair divergence functions because they are not positive-definite.
However, the analysis of \cite{ciaglia_dicosmo_felice_mancini_marmo_perez-pardo-hamilton-jacobi_approach_to_potential_functions_in_information_geometry} clearly shows that we may add terms of at least fourth order in the velocities to $\lag_{\alpha}$ and the resulting Hamilton principal function will be again a potential function for $(\mathcal{M}\,,g\,,T)$.
Consequently, we could keep adding terms of higher order in the velocities so that the resulting potential function is actually a divergence function.
 
In this short contribution we want to formulate an inverse problem for the Hamilton-Jacobi theory focused on some relevant situations in information geometry.
Specifically, we ask the following question: Given a fixed divergence function $D$ on $\mathcal{M}\times \mathcal{M}$ generating the statistical model $(\mathcal{M}\,,g\,,T)$, is it possible to find a Lagrangian function $\lag$ on $T\mathcal{M}$ such that $D$ is the Hamilton principal function $S$ of $\lag$?
If the answer is yes, then we can analyze the associated dynamical system and its physical interpretation in the context of the adopted model.
In the following we will review a case in which the answer exist in full generality, namely, the case of  of self-dual statistical manifolds (\cite{amari_nagaoka-methods_of_information_geometry}).
An interesting example of such a manifold is given by the space of pure states of quantum mechanics which will be briefly discussed.
The possibility to extend this ideas to relevant cases going beyond self-dual statistical manifolds will be addressed in future works.

\section{Hamilton-Jacobi, information geometry, and the inverse problem for potential functions}

Self-dual statistical manifolds (see \cite{amari_nagaoka-methods_of_information_geometry}) are statistical models for which the symmetric tensor $T$ identically vanishes, so that the only connection available is the self-dual Levi-Civita connection $\nabla_{g}$ associated with the metric $g$, and a canonical contrast function $D_{d}$ is given by:
\be\label{eqn: contrast function for self-dual manifolds}
D_{d}(m_{\mathrm{in}}\,,m_{\mathrm{fin}})=\frac{1}{2}\,d^{2}(m_{\mathrm{in}}\,,m_{\mathrm{fin}})\,,
\ee
where $d^{2}(m_{\mathrm{in}}\,,m_{\mathrm{fin}})$ is the square of the Riemannian geodesic distance associated with the metric $g$ on $\mathcal{M}$.
In this particular case, it turns out (see \cite{ciaglia_dicosmo_felice_mancini_marmo_perez-pardo-hamilton-jacobi_approach_to_potential_functions_in_information_geometry}) that the family $\lag_{\alpha}$ of Lagrangian functions given in equation \eqref{Lagrangian} provides a solution to the inverse problem.
Indeed, when $T=0$, the family of Lagrangian functions $\lag_{\alpha}$ collapses to a single Lagrangian  which is  the metric Lagrangian $\lag_{g}=\dfrac{1}{2} g_{jk}v^jv^k$.
To prove that $\lag_{g}$ actually solves the inverse problem for $S_{d}$ in the case of self-dual manifolds, let us recall that, if the manifold $\mathcal{M}$ is regular enough, the square of Riemannian geodesic distance $d^{2}(m_{\mathrm{in}}\,,m_{\mathrm{fin}})$ is given by:
\be\label{eqn: riemannian distance}
d^{2}(m_{\mathrm{in}}\,,m_{\mathrm{fin}})=\left(\int_{0}^{1}\,\sqrt{g_{jk}(\gamma_{g}(t))\,\dot{\gamma_{g}}^{j}\,\dot{\gamma_{g}}^{k}}\,\mathrm{d}t\right)^{2}=\left(\int_{0}^{1}\,\sqrt{2\lag_{g}(\gamma_{g},\dot{\gamma_{g}})}\,\mathrm{d}t\right)^{2}
\ee
where $\gamma_{g}$ is a geodesics for $g$ with fixed endpoints $m_{\mathrm{in}}$ and $m_{\mathrm{fin}}$, and where the square root is introduced in order to ensure the invariance of the distance function under reparametrizations of $\gamma$.
Geodesics curves are precisely the projection of the integral curves of a vector field $\Gamma$  on $T\mathcal{M}$ which is the dynamical vector field associated with the Lagrangian function $\lag_{g}=\frac{1}{2}g_{jk}\,v^{j}\,v^{k}$  by means of the Euler-Lagrange equations stemming from the variational principle for the action functional \eqref{eqn: action}.
Now, recall that the metric Lagrangian $\lag_{g}$, as well as all of its functions $F(\lag_{g})$ with $F$ analytic, give rise to the same dynamical trajectories (\cite{marmo_ferrario_lovecchio_morandi_rubano-the_inverse_problem_in_the_calculus_of_variations_and_the_geometry_of_the_tangent_bundle}) and   are all constants of the motion for this dynamics.
Consequently, we can take $\sqrt{2\lag_{g}(\gamma_{g},\dot{\gamma_{g}})}$ out of the integral in equation \eqref{eqn: riemannian distance} so that we are left with:
$$
D_{d}(m_{\mathrm{in}}\,,m_{\mathrm{fin}})=\frac{1}{2}d^{2}(m_{\mathrm{in}}\,,m_{\mathrm{fin}})=\lag_{g}(\gamma_{g},\dot{\gamma_{g}})=
$$
\be
=\int_{0}^{1}\,\lag_{g}(\gamma_{g}\,,\dot{\gamma_{g}})\,\mathrm{d}t=I(\gamma_{g})=S(m_{\mathrm{in}}\,,m_{\mathrm{fin}})\,,
\ee
which means that the canonical divergence function of self-dual manifolds is actually  the Hamilton principal function of the metric Lagrangian $\lag_{g}$ associated with the metric tensor $g$.

A relevant example of self-dual manifold is given by the space $\mathcal{P}(\mathcal{H})$ of pure states of a quantum system with Hilbert space $\mathcal{H}$.
We are here considering probability amplitudes instead of probability distributions.
For simplicity, we limit our case to the finite-dimensional case $\mathcal{H}\cong\mathbb{C}^{n}$.
The metric $g$ on $\mathcal{P}(\mathcal{H})$ is the so-called Fubini-Study metric   (see \cite{carinena_ibort_marmo_morandi-geometry_from_dynamics_classical_and_quantum}).
Apart from a constant conformal factor, $g$  is the unique  metric on $\mathcal{P}(\mathcal{H})$ which is invariant under the canonical action of the unitary group $\mathcal{U}(n)$ on $\mathcal{P}(\mathcal{H})$.
The manifold $\mathcal{P}(\mathcal{H})$ is a homogeneous space for the unitary group, specifically, it is $\mathcal{P}(\mathcal{H})\cong\mathcal{U}(n)/\mathcal{U}_{\rho_{\psi}}$, where $\mathcal{U}_{\rho_{\psi}}$ is the istropy subgroup of the non-negative Hermitean matrix $\rho_{\psi}$ associated with a pure state $\psi$ with respect to the action $\rho_{\psi}\mapsto U^{\dagger}\,\rho_{\psi}\,U$ for which the space of pure states is a homogeneous space of the unitary group.
Note that $(\mathcal{P}(\mathcal{H})\,,g)$ is a Riemannian homogeneous manifold.
We may exploit the homogeneous space structure of $\mathcal{P}(\mathcal{H})$ in order to describe the Lagrangian function associated with the metric tensor $g$ by means of a degenerate Lagrangian function on the tangent bundle of the unitary group.
This is particularly useful since $\mathcal{U}(n)$ is a Lie group, hence it is parallelizable, and thus a pair of global dual basis $\{X_{j}\}$ and $\{\theta^{j}\}$ of, respectively, vector fields and one-forms are available.
Let us consider then a fixed  positive matrix $\rho_{\psi}$ associated with a fiducial pure state $\psi$, and consider the  following Lagrangian:
\be
\lag(\gr\,,\grd)=\frac{1}{2}Tr\left(\left[\rho_{\psi}\,,\gri\grd\right]^{2}\right)=\frac{1}{2}G_{jk}\dot{\theta}^{j}\dot{\theta}^{k}\,,
\ee
where $G_{jk}$ is a constant matrix, and $\dot{\theta}^{j}$ is the velocity-like function defined on the tangent space of every Lie group (see \cite{marmo_rubano-particle_dynamics_on_fiber_bundles}).
It is clear that $\lag$ is invariant with respect to the tangent lift of the left action of $\mathcal{U}(n)$ on itself and with respect to the tangent lift of the right action of the isotropy subgroup $\mathcal{U}_{\rho_{\psi}}$.
Consequently, $\lag$ is the pullback to $T\mathcal{U}(n)$ of a Lagrangian function $\mathcal{L}$ on $\mathcal{P}(\mathcal{H})$.
In order to focus on the main stream of the paper, we will not enter into a full discussion for this dynamical system.
We simply state that, using the theory of degenerate Lagrangians (\cite{marmo_mukunda_samuel-dynamics_and_symmetry_for_constrained_systems_a_geometrical_analysis}), it is possible to prove that $\mathcal{L}$ is the metric Lagrangian associated with the Fubini-Study metric, and that the dynamical trajectories of the vector field $\Gamma$ associated with $\lag$ project down onto the geodesics of the Fubini-Study metric on the space of quantum pure states.
Specifically, writing $\rho_{0}=U_{0}^{\dagger}\,\rho_{\psi}\,U_{0}$, we have:
\be
\gamma_{\rho_{0},\mathbf{A}}(t)=\mathrm{e}^{- \left[\rho_{\psi}\,,\mathbf{A}\right]t}\,\rho_{0}\,\mathrm{e}^{\left[\rho_{\psi}\,,\mathbf{A}\right]t}\,,
\ee
where $\mathbf{A}$ is a self-adjoint matrix.
The dynamical vector field on $T\mathcal{P}(\mathcal{H})$ may be seen as a family of vector fields on $\mathcal{P}(\mathcal{H})$ labelled by the matrix parameter $\mathbf{A}$.
Once we select a member of this family, that is we fix $\mathbf{A}$, we are left with a vector field on the space of pure quantum state  generating the unitary evolution associated with the Hamiltonian operator $\mathbf{H}=-\imath[\rho_{\psi}\,,\mathbf{A}]$.
These evolutions have a clear physical meaning, indeed, they represent the dynamical evolution of an isolated quantum system with energy operator $\mathbf{H}$.
The Hamilton principal function for $\lag$ is the pullback of the Hamilton principal function associated with the Lagrangian function $\mathcal{L}$ on $T\mathcal{P}(\mathcal{H})$.
Writing $\rho_{1}=\gamma_{\rho_{0},\mathbf{A}}(1)$ we have:
\be
S(\rho_{0}\,,\rho_{1})=\frac{1}{2}Tr\left(\left[\rho_{\psi}\,,\left[\rho_{\psi}\,,\mathbf{A}\right]\right]^{2}\right)=Tr\left(\rho_{\psi}\,\mathbf{A}\,\left[\mathbf{A}\,,\rho_{\psi}\right]\right)\,.
\ee
For example, let us consider a two-level quantum system, for which the most general pure state is:
\be
\rho=\frac{1}{2}\left(\mathbb{I} + x^{j}\sigma_{j}\right)\,,
\ee
where $\mathbb{I}$ is the identity matrix, the $\sigma_{j}$'s are the Pauli matrices, and $\delta_{jk}x^{j}x^{k}=1$.
We take $\rho_{\psi}=\frac{1}{2}\left(\mathbb{I} + \sigma_{3}\right)$.
In this case, the isotropy subgroup $\mathcal{U}(2)_{\rho_{\psi}}$ is equal to $\mathcal{U}(1)\times\mathcal{U}(1)$,  and thus $\mathcal{P}(\mathcal{H})$ is a two-dimensional sphere embedded in the three-dimensional space $\mathbb{R}^{3}$.
The tensor $g$ reads:
\be
g=G_{nk}\theta^{n}\otimes\theta^{k}=\theta^{1}\otimes \theta^{1} + \theta^{2}\oplus\theta^{2}\,.
\ee
A direct computation shows that the dynamical trajectories are:
\be
\rho(t)=\cos(rt)\rho_{0} + \frac{\sin(rt)}{r}[\rho_{\psi}\,,\mathbf{A}]\,,
\ee
where, $r^{2}=(A^{1})^{2} + (A^{2})^{2}$. 
From this it follows that:
\be
[\rho\,,\mathbf{A}]=\frac{1}{\sqrt{1-(\delta_{jk}\,x_{0}^{j}\,x_{1}^{k})^{2}}}\arccos\left(\delta_{jk}\,x_{0}^{j}\,x_{1}^{k}\right)\,[\rho_{0}\,,\rho_{1}]\,,
\ee
and thus:
\be
S(\rho_{0}\,,\rho_{1})=\arccos^{2}\left(\delta_{jk}\,x_{0}^{j}\,x_{1}^{k}\right)\,.
\ee

Going back to probability distributions, let us recall a particular case in which the inverse problem formulated here has a positive solution (see \cite{ciaglia_dicosmo_felice_mancini_marmo_perez-pardo-hamilton-jacobi_approach_to_potential_functions_in_information_geometry}).
Consider the following family of exponential distributions on $\mathbb{R}^{+}$ parametrized by $\xi\in\mathbb{R}^{+}=\mathcal{M}$:
\be
\mathit{p}(x\,,\xi)=\xi\,\mathrm{e}^{-x \xi}\qquad \xi,x>0\,.
\ee
The Kullback-Leibler divergence function for this model is:

\be\label{eqn: K-L divergence 1d example}
D_{KL}(\xi_{\mathrm{in}}\,,\xi_{\mathrm{fin}})=\int_{0}^{+\infty}\,\mathit{p}(x\,,\xi_{\mathrm{in}})\ln\left(\frac{\mathit{p}(x\,,\xi_{\mathrm{in}})}{\mathit{p}(x\,,\xi_{\mathrm{fin}})}\right)\mathrm{d}x=\ln\left(\frac{\xi_{\mathrm{in}}}{\xi_{\mathrm{fin}}}\right) + \frac{\xi_{\mathrm{fin}}}{\xi_{\mathrm{in}}} - 1\,.
\ee
A direct computation shows that $D_{KL}$ is the Hamilton principal function associated with the Lagrangian function:
\be
\lag_{KL}(\xi\,,v)= \mathrm{e}^{\frac{v}{\xi}} - \frac{v}{\xi} - 1\,.
\ee
In this case, it happens that the dynamical system associated with $\lag_{KL}$ and the dynamical system associated with the metric Lagrangian $\lag_{g}$ of this statistical model are the same, that is, $\lag_{KL}$ and $\lag_{g}$ are non gauge-equivalent alternative Lagrangians (see \cite{marmo_ferrario_lovecchio_morandi_rubano-the_inverse_problem_in_the_calculus_of_variations_and_the_geometry_of_the_tangent_bundle}).

\section{Conclusions}

We have seen how the inverse problem for divergence functions in the context of Hamilton-Jacobi theory has a positive answer in the case of self-dual statistical manifolds.
In this case, the canonical divergence function $D(m_{1}\,,m_{2})=\frac{1}{2}d^{2}(m_{1}\,,m_{2})$, where $d^{2}(m_{1}\,,m_{2})$ is the Riemannian distance, is the Hamilton principal function associated with the metric Lagrangian $\lag_{g}$.
In the case when $\mathcal{M}$ is the space of pure states of a finite-level quantum system, the metric $g$ is the Fubini-study metric and we have seen how to describe the metric Lagrangian $\lag_{g}$ by means of a degenerate Lagrangian $\lag$ on the unitary group.

In general, both in classical and quantum information geometry, some well-known divergence functions are  relative entropies (see \cite{amari_nagaoka-methods_of_information_geometry,balian-the_entropy_based_quantum_metric,balian_alhassid_reinhardt-dissipation_in_many-body_systems_a_geometric_approach_based_on_information_theory,manko_marmo_ventriglia_vitale-metric_on_the_space_of-quantum_states_from_relative_entropy_tomographic_reconstruction,petz-quantum_information_theory_and_quantum_statistics}), hence, a positive answer to the inverse problem for such divergence functions brings in the possibility of defining dynamical systems associated with relative entropies, and, in accordance with the Hamilton-Jacobi theory, this points to the possibility of looking at relative entropies as generators of canonical transformations.
A more thourogh analysis of these situations will be presented in future works.

Finally, let us comment on the possible relation of this work with the recent developments in Souriau's Lie group thermodynamic.
In this framework, a sort of Hessian metric, called Souriau-Fisher metric, $g$ is defined on a manifold $\mathcal{M}$ by means of a function on $\mathcal{M}$, the so-called Koszul-Vinberg Characteristic function (see \cite{barbaresco-geometric_theory_of_heat_from_Souriau_lie_groups_thermodynamics_and_koszul_hessian_geometry_applications_in_information_geometry_for_exponential_families,marle-from_tools_in_symplectic_and_poisson_geometry_to_souriau’s_theories_of_statistical_mechanics_and_thermodynamics}) .
It is not possible to compare directly our procedure with the Koszul-Vinberg Characteristic function generating the same statistical structure since the latter is a function defined on $\mathcal{M}$ and not on $\mathcal{M}\times \mathcal{M}$.
Moreover, one has to generalize the Hamilton-Jacobi approach along the lines explained in \cite{marmo_morandi_mukunda-a_geometrical_approach_to_the_hamilton-jacobi_form_of_dynamics_and_its_generalizations}, section 6. 
This generalization amounts to replace $\mathbb{R}$ of the extended formalism with a Lie Group (which could be the Galilei group or Poincar\`e group).
The nontriviality of the second cohomology group for the Galilei group would require to work with suitable central extensions to apply the generalized theory. 
In Souriau's theory, the so-called Euler-Poincar\`e equations naturally appear.
These equations are equivalent to the equations of motion of a Lagrangian system with symmetries, however, they are defined on the product of the configuration space with the Lie algebra of the group of symmetries of the system rather than on the tangent bundle of the configuration space.
Furthermore, they may be derived starting from a variational principle just like Euler-Lagrange equations.
Consequently, a possible relation between Hamilton principal function for the action \eqref{eqn: action} and the Koszul-Vinberg Characteristic function will be possible when the Hamilton-Jacobi theory is generalized to include a Lie group $G$ instead of $\mathbb{R}$.

\section{Acknowledgement}

G.M. would like to acknowledge the partial support by the
‘‘Excellence Chair Program, Santander-UCIIIM’’

\end{document}